# Resonant nonlinear refraction of 4 – 5 μm light in CO and $CO_2$ gas

J.J. Pigeon[1], D. Tovey[2], S. Ya. Tochitsky[2], G. J. Louwrens[2], I. Ben-Zvi[1]
D. Martyshkin[3], V. Fedorov[3], K. Karki[3], S. Mirov[3], and C. Joshi[2]

[1]*Department of Physics and Astronomy, Stony Brook University, Stony Brook, New York, 11794, USA*
[2]*Department of Electrical Engineering, University of California at Los Angeles, Los Angeles, California 90095, USA*
[3]*Department of Physics, University of Alabama at Birmingham, Birmingham, Alabama 35294, USA*
jeremy.pigeon@stonybrook.edu

## Abstract

The resonant nonlinear refraction of 4 – 5 μm light in CO and $CO_2$ gas at a peak intensity of 15 MW/cm$^2$ was demonstrated using time- and frequency-resolved measurements of self-focusing and self-defocusing. The nonlinearity of these molecular gases exhibits intensity-dependent sign reversals and a < 4 ns response time. A change from self-focusing to self-defocusing or vice-versa was observed to occur for Rabi frequencies that are comparable to the collisional linewidth. A density matrix model for the nonlinear susceptibility of a strongly driven two-level system provides a qualitative explanation for these results.

## I. Introduction

Resonant nonlinear optics provides a means to measure properties of physical systems and to study light-matter interaction. Propagation in a nonlinear medium changes various parameters of a driving laser pulse that can be measured to identify the quantum pathways that contribute to the nonlinearity and gain insight into otherwise inaccessible material parameters. Experiments on the resonant nonlinear optics of simple systems, such as atomic and molecular gases, have been instrumental in furthering understanding of optical physics. Experiments in this field have lead to the discovery of many important effects, such as Autler-Townes splitting [1], resonant nonlinear refraction [2, 3], electromagnetic induced transparency [4], and the production of slow light [5]. Aside from fundamental importance, many of these effects find practical application in the development of lasers and in quantum optics [6].

Most experiments on the resonant nonlinear optics of gases have involved visible or near-IR light interacting with the electronic transitions of atomic gases. In atomic gases, the transition dipole moment, μ, is approximately 5 D and resonant nonlinear refraction leading to self-focusing or self-defocusing is dominated by saturable absorption [7]. The nonlinearity associated with the ~ 0.3 D, rovibrational transitions of simple molecules has been much less studied due to a lack of high-power sources. For experiments in molecular gases, the strong, $μ^4$, scaling of the nonlinear refractive index with dipole moment necessitates the use of high-power sources of radiation at wavelengths that are challenging to produce using solid-state lasers. As a result, previous experiments on the resonant nonlinear optics of molecules have been performed using line-tunable $CO_2$ lasers having wavelengths near 9 – 10 μm [*e.g.*, 8 – 12]. These experiments were constrained to the coincidental overlap of $CO_2$ laser lines with the absorption lines of molecules, thereby limiting experiments to molecules having complex absorption spectra such as $NH_3$, $SF_6$, and $CDF_4$.

The advent of broadband, transition metal doped chalcogenide lasers such as Cr:ZnSe and Fe:ZnSe, have made continuously tunable sources of high-power, mid-IR radiation available [13, 14]. We have recently used a nanosecond-class, 4 – 5 µm, Fe:ZnSe laser, for the first experiment on resonant nonlinear refraction of 4.3 µm radiation in $CO_2$ gas [15]. This experiment has revealed that the nonlinear optics of $CO_2$ gas cannot be explained by the saturable absorption model [7] that is widely applicable to atomic gases or more complex molecules. Rather, we have discovered that the resonant nonlinearity of $CO_2$ appears to be dominated by field effects such as power-broadening and ac-Stark splitting instead of saturable absorption [15]. This manifested in a resonant nonlinear refractive index that was opposite in sign and a response time that was fast compared to a system dominated by saturable absorption. Measurements have also indicated that the resonant nonlinearity of 100 torr of $CO_2$ is ~ $10^{-12}$ $cm^2/W$, a nonlinearity comparable with n-Ge in this wavelength range. This large optical nonlinearity suggests that the ambient $CO_2$ concentration in the air may decrease the critical power for self-focusing for relatively narrowband radiation by $10^4$ times as compared to that measured using broadband LWIR and mid-IR laser pules [16 – 19].

In this article, we report detailed measurements of resonant nonlinear refraction of 4 – 5 µm radiation in CO and $CO_2$ gas. By studying self-focusing and self-defocusing as a function of frequency detuning, we have found that both CO and $CO_2$ gas exhibit a fast nonlinear response that is strongly affected by field effects. Whereas previously [15] we have observed that the sign of the resonant nonlinear refractive index of $CO_2$ was always opposite that of a system dominated by saturable absorption, measurements of resonant nonlinear refraction in CO and $CO_2$ over a range of intensities in the ~$10^7$ $W/cm^2$ range have revealed that the resonant nonlinear optics of these simple molecules are much more complicated. We have observed that the sign of the nonlinearity of these molecules can reverse sign from self-focusing to self-defocusing or vice-versa when the Rabi frequency becomes comparable to the collisional linewidth. We have compared the experimental results to calculations of the electric susceptibility using a simplified density matrix model that suggests that the observed behavior is qualitatively consistent with that of a strongly driven two-level system. These results have implications for the propagation of high-power, infrared radiation in the atmosphere and nonlinear pulse propagation [20].

## II. Experimental method

Measurements of resonant nonlinear refraction of 4 – 5 µm radiation in CO and $CO_2$ gas were accomplished using a continuously tunable Fe:ZnSe laser as a pump source. The room temperature Fe:ZnSe laser was pumped by 2.94-µm, 15 mJ, laser pulses produced by a Q-switched Er:YAG oscillator to produce 40 ns, 2 mJ, 3.6 – 5.1 µm laser pulses at 10 Hz [14]. Figure 1a, below, shows a simplified experimental set-up used for measurements of resonant nonlinear refraction. We focused ~ 2 mJ pulses using two different beam focusing geometries, one provided a peak intensity ~ 15 $MW/cm^2$ and the second with a peak intensity of ~ 7.5 MW/cm$^2$. In either case, the Rayleigh length was longer than the 10 cm long, gas-filled cell used for measurements. After the cell, a 50/50 beam splitter was used to realize simultaneous temporal and spatial or spectral measurements. Time-resolved measurements are accomplished using a fast (~ 1ns temporal resolution) HgCdTe detector with a small aperture (1 mm$^2$). We imaged the exit of the cell with 5x magnification to accomplish spatial measurements using a pyroelectric camera. Spectrally-resolved measurements were performed using a spectrometer with ~ 1 nm resolution and a pyroelectric camera as a read-out device.

The inset of Fig. 1 shows simplified energy level diagrams of CO and the asymmetric stretching mode of $CO_2$. The approximate dipole moments of the transitions under investigation are approximately 0.3 and 0.1 D for $CO_2$ and CO, respectively [21, 22]. We have observed both self-focusing and self-defocusing in $CO_2$ but only self-defocusing in CO, owing to the smaller dipole moment. Figure 1b and 1c show the calculated absorption spectra of $CO_2$ and CO, respectively. We have performed measurements in the vicinity of multiple lines belonging the 4P and 4R branch of the 000 – 001 transition of $CO_2$ and the 4P branch of the $\upsilon = 0$ to $\upsilon = 1$ transition of CO. The regions where spectral measurements were recorded are marked by the shaded rectangles on Fig. 1b and Fig. 1c. It should be noted that the rotational constant, B, of CO is approximately three times larger than that of $CO_2$ resulting in a larger line spacing as shown in Fig. 1c. Measurements in $CO_2$ were accomplished using 50 – 100 torr of pure $CO_2$ while measurements in CO were accomplished using 300 – 760 torr of 1:9 CO:He mix. The dilute CO mix was used only for safety purposes. It should be noted that the bandwidth of the laser was approximately 2 nm FWHM, which is 10 – 100x broader than the collisional linewidth of the gases under investigation. Measurements in the vicinity of a single line were still possible, however, as the separation between lines for the 4P branch of $CO_2$ is ~ 4 nm and the line separation for CO is ~12 nm (see Fig. 1b and 1c).

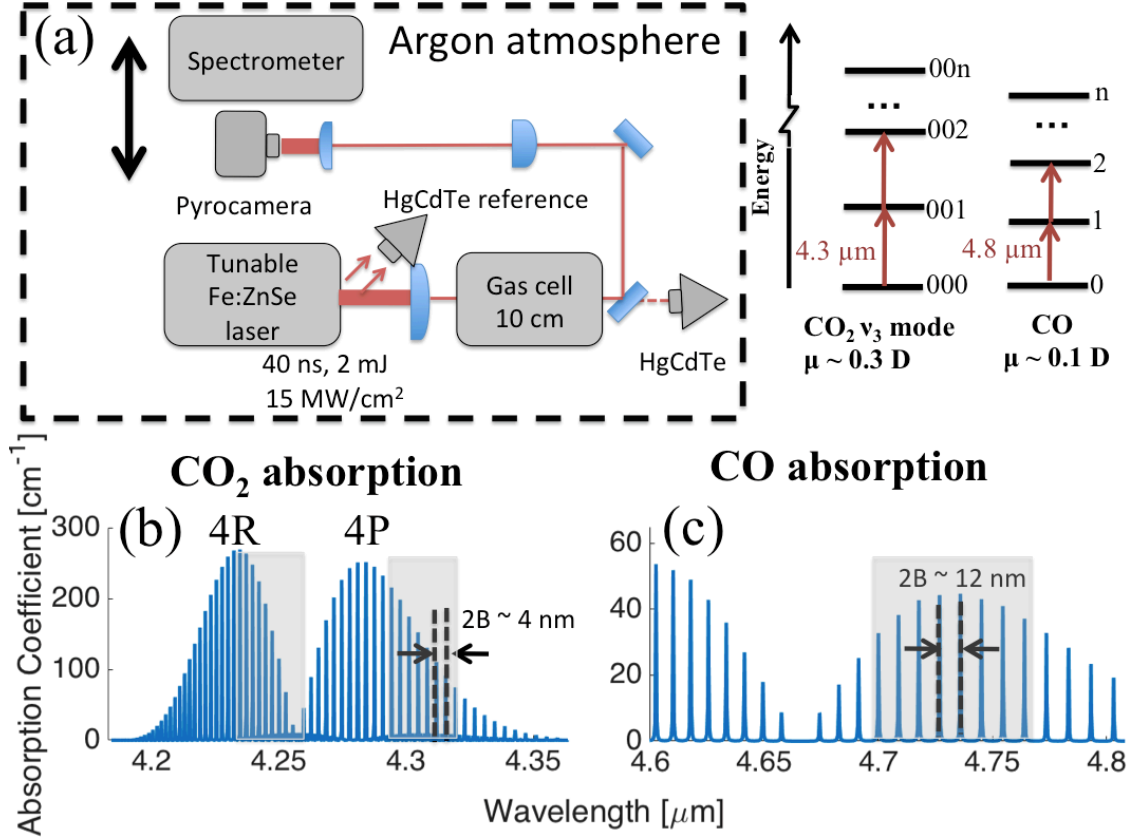

Figure 1: (a) The simplified experimental set-up used for time- and frequency-resolved measurements of resonant nonlinear refraction of 4 – 5 µm light in CO and $CO_2$ gas. (b) The absorption spectrum of 100 torr of $CO_2$ gas in the 4 µm wavelength range calculated using HITRAN [25]. (c) The absorption spectrum of 1 atm of CO gas calculated using HITRAN. The inset is a simplified energy level diagram of CO and $CO_2$. Regions where spectral measurements were recorded are marked by the shaded rectangles on (b) and (c).

### III. Measurements of resonant nonlinear refraction in $CO_2$

We have previously observed self-focusing, self-defocusing, and spatial beam break-up indicative of multiple filamentation in ~ 100 torr of $CO_2$ gas pumped near 4.3 µm [15]. The onset of beam breakup and a Gaussian beam decomposition method [12, 26] were used to determine that the nonlinear refractive index of 100 torr of $CO_2$ is ~ $10^{-12}$ $cm^2$/W [15], a nonlinear refractive index that is remarkably comparable to semiconductors such as n-Ge. Frequency-resolved measurements of resonant refraction in the vicinity of the 4P(30) line are depicted in Fig. 2, where the vertical axis represents the spatial extent of the laser beam and the horizontal axis represents wavelength. These frequency-resolved measurements, obtained using 50 and 100 torr of $CO_2$ (Fig. 2a and Fig. 2b, respectively) excited with a peak intensity of ~ 15 MW/$cm^2$, indicate self-focusing and self-defocusing of the laser spectrum on the red and blue sides of resonance, respectively. The dumbbell shape shown in Fig. 2a and 2b is a vertical cross-section of the donut-shaped beam that is observed in a case of strong self-defocusing. A similar donut-shaped beam has also been reported in experiments on resonant nonlinear refraction in atomic vapors [3].

Frequency-resolved measurements of nonlinear refraction show features consistent with the power-broadening of a rovibrational line. A ring of attenuation caused by this effect is visible in the frequency-resolved measurements depicted in Fig. 2. Power-broadening causes the rovibrational line to broaden in proportion to the resonant Rabi frequency, $\upsilon_{Rabi} = \mu E/h$ where $\mu$ is the transition dipole moment and E is the peak amplitude of the electric field. The amount of broadening is largest at the beam center where the intensity of the Gaussian beam is maximum resulting in the ring shape visible in Fig. 2. Figure 2a and 2b were obtained in 50 and 100 torr of $CO_2$, respectively. The use of a higher pressure results in a larger diameter ring in Fig. 2b due to the collisional broadening of the transition. Using the 15 MW/cm$^2$ peak intensity and the 0.3 D dipole moment of these $CO_2$ transitions [21], we estimate that the Rabi frequency is ~ 15 GHz. This value is an overestimate of the Rabi frequency, however, since the bandwidth of the laser is ~ 100x broader than the 100 torr collisional linewidth and only a fraction of the laser intensity interacts with the transition. To meaningfully compare experimental results obtained at various pressures, we have accounted for this effect by estimating the effective Rabi frequency as $\upsilon_{Rabi} = \mu E/h (\Delta\upsilon/\Delta\upsilon_L)^{1/2}$ where $\Delta\upsilon_L$ is the laser bandwidth. With these considerations, we estimate an effective Rabi frequency in the few GHz range. Nevertheless, this effective Rabi frequency was still larger than the ~500 MHz collisional linewidth of the $CO_2$ absorption lines at 100 torr [28], resulting in a normalized Rabi frequency, $\upsilon_{Rabi} \Delta\upsilon^{-1}$, of ~ 3 as indicated on Fig. 2b.

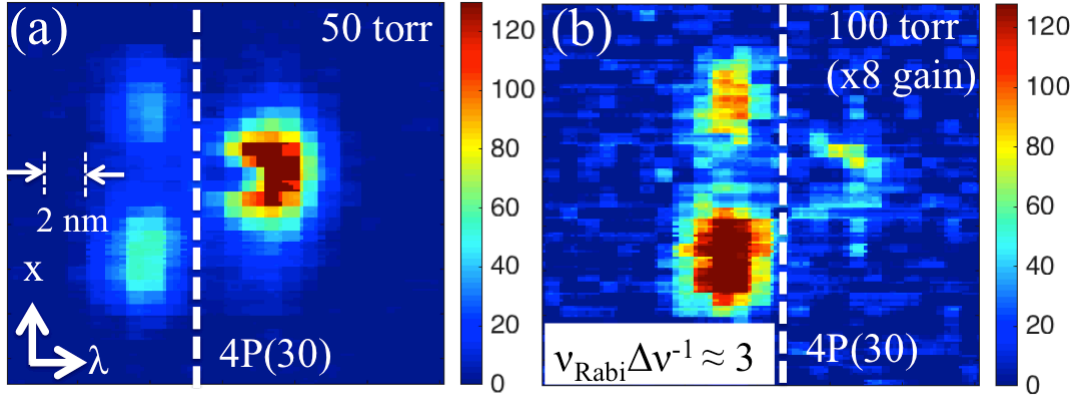

Figure 2: Frequency-resolved measurements of self-focusing and self-defocusing in the vicinity of the 4P(30) line of the 000 – 001 transition of $CO_2$ using (a) 50 torr and (b) 100 torr of $CO_2$.

By increasing the beam size the peak intensity was decreased from ~ 15 MW/cm$^2$ to ~7.5 MW/cm$^2$. By doing so, we have observed a reversal of the sign of the resonant nonlinearity of $CO_2$ gas. Figure 3, below, shows frequency-resolved refraction measurements that depict a transition from self-focusing to self-defocusing for a case when the laser was tuned to the red-side of the 4P(36) line of the 000 – 001 transition of $CO_2$ having a pressure of 60 torr. Figure 3a, obtained using 2 mJ (~7.5 MW/cm$^2$) 4.3 μm pulses indicates a self-focusing nonlinearity on the red side of resonance while Fig. 3b, obtained using 1 mJ pulses (~4 MW/cm$^2$) shows a self-defocusing nonlinearity on the red side of resonance. By decreasing the intensity we have changed the ratio of the effective Rabi frequency to the collisional linewidth from $\upsilon_{Rabi} \Delta\upsilon^{-1}$ ~ 1.5 (Fig. 3a) to $\upsilon_{Rabi} \Delta\upsilon^{-1}$ ~1 (Fig. 3b) as indicated in Fig. 3. The observation of the sign reversal at Rabi frequencies approximately equal to the collisional linewidth supports the hypothesis that field effects such as power broadening and ac-Stark splitting strongly affect the nonlinearity of $CO_2$.

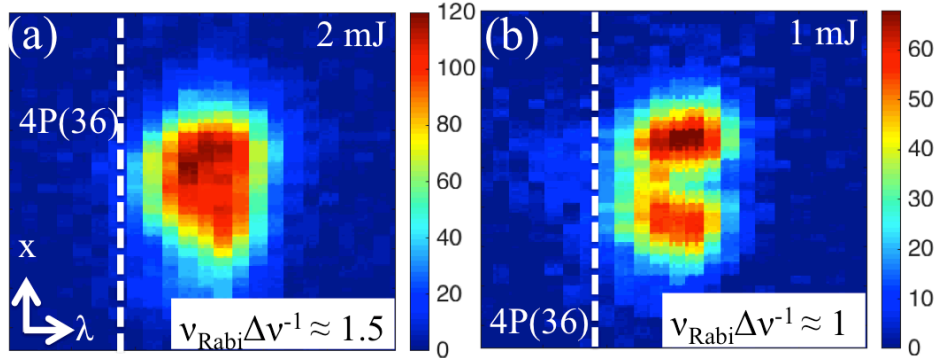

Figure 3: Frequency-resolved measurements of self-focusing and self-defocusing on the red side of the 4P(36) line of the 000 – 001 transition of $CO_2$ for (a) a 2 mJ pump and a (b) 1 mJ pump, showing an intensity-dependent sign reversal observed in 60 torr of $CO_2$. These measurements were performed with approximately twice the cross-sectional area as those presented above to enable peak intensities of ~ 4 – 8 MW/cm$^2$. Approximate normalized Rabi frequencies ($\nu_{Rabi} \Delta\nu^{-1}$) are indicated on each figure.

## IV. Measurements of resonant nonlinear refraction in CO

Due to the strong scaling of the resonant nonlinear refractive index with the transition dipole moment, $n_2 \sim \mu^4$, we have only observed self-defocusing in CO gas having a 3x smaller dipole moment than $CO_2$ [21, 22]. Figure 4a and 4b show spatial beam profiles of the initial 4.8 μm beam and the beam after it has propagated through 300 torr of 1:9 CO:He mix, respectively. For these measurements, we have used a peak intensity of ~ 8 MW/cm$^2$ with the laser spectrum tuned to the red side of the 4P(8) line. Note that the 4P(8) line of the $\upsilon = 0$ to $\upsilon = 1$ band is located near the peak of the 4P branch at room temperature (see Fig. 1). The spatial beam profile presented in Fig. 4b does not clearly depict self-defocusing since narrowing of the beam caused by strong absorption obscures this effect. The best indication of nonlinear refraction was the transformation of the initially elliptical beam to a dumbbell shape as can be seen in Fig 4b. In this case, self-defocusing may result in a dumbbell as opposed to the typically observed ring shape due to the asymmetric initial beam profile.

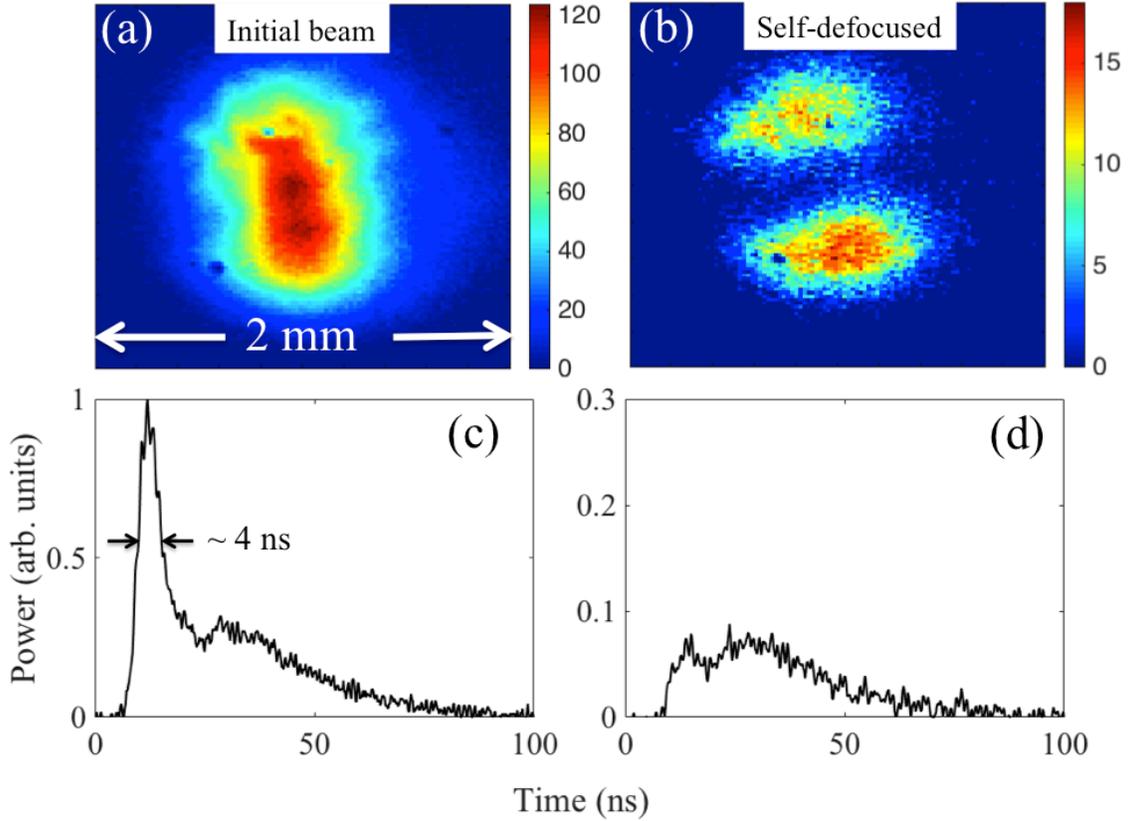

Figure 4: Beam profile measurements of (a) the initial beam and (b) self-defocused beam that correspond to temporal pulse profile measurements (c) and (d), respectively. Note that pulse consisted of a 4 ns FWHM spike followed by a 30 ns tail and that only the initial spike was subjected to nonlinear focusing. The vertical axes of (c) and (d) are normalized to the peak amplitude of the initial laser pulse. These measurements were performed in the vicinity of the 4P(8) line of the $\upsilon = 0$ to $\upsilon = 1$ transition of CO for which the absorption was ~50%.

Time-resolved measurements of nonlinear refraction indicated clear self-defocusing with a nonlinear response time ≤ 4 ns, a similar response time to that measured for $CO_2$ previously [15]. Figure 4a and 4c show the initial spatial beam profile and temporal pulse profile, respectively, while Fig. 4b and 4d show the spatial and temporal profiles after the beam has been self-defocused by 500 torr of 1:9 CO:He mix. As can be seen from Fig. 4d, only the leading 4 ns spike was subjected to the self-defocusing effect suggesting that the response time is ≤ 4 ns.

Frequency-resolved measurements of nonlinear refraction in CO gas indicate that the sign of the nonlinearity can also change depending on the laser intensity and gas pressure, as was the case with $CO_2$. Figure 5, below, shows frequency-resolved measurements of nonlinear refraction in 1 atm of 1:9 CO:He mix in the vicinity of the 4P(8) transition of the $\upsilon = 0$ to $\upsilon = 1$ transition of CO. For a pressure of 1 atm and an intensity of 8 MW/cm$^2$, we estimate that the collisional linewidth is ~ 1.5 GHz [29] and the effective Rabi frequency is ~ 1 GHz ($\upsilon_{Rabi}\, \Delta\upsilon^{-1}$ ~ 0.7). As in the case of $CO_2$, the effective Rabi frequency was reduced since the laser bandwidth is ~ 20x larger than the collisional linewidth and only a fraction of the laser power interacts with the transition. For these parameters, the beam was self-defocused on the red side of the 4P(8) transition (see Fig. 5b). Although the sign of the nonlinearity was typically consistent with that described above, we have also observed a sign reversal in 300 torr of 1:9 CO:He mix when using the beam geometry with a 2x smaller cross-sectional area. For these conditions we estimate that the collisional linewidth is ~600 MHz [29] and effective Rabi frequency near 700 MHz ($\upsilon_{Rabi} \Delta\upsilon^{-1} \sim 1.2$). These observations suggest that the resonant nonlinearity of CO is also strongly affected by field effects.

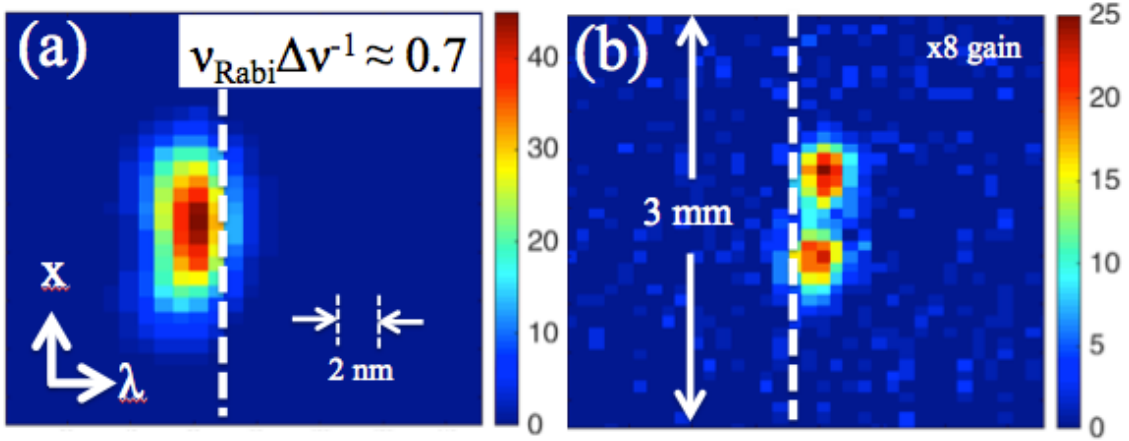

Figure 5: Frequency-resolved measurements in the vicinity of the 4P(8) line of the $\upsilon = 0$ to $\upsilon = 1$ transition of CO. (a) shows the unaffected beam profile and (b) shows strong self-defocusing red side of the rovibrational line. These measurements were obtained using 1 atm of 1:9 CO:He mix at a peak intensity of ~ 8 MW/cm$^2$.

## V. Rabi frequency dependence of the sign of the resonant nonlinearity of CO and CO$_2$ gas

In [15] we have presented a simplified physical picture to explain the resonant nonlinear refraction of 4.3 μm radiation in CO$_2$ gas. This explanation is based on the hypothesis that the ac-Stark effect causes the absorption line to split at Rabi frequencies larger than the collisional linewidth resulting in a sign reversal in the resonant nonlinear refractive index. In this experiment, the range of pressures and laser intensities used for measurements in both CO and CO$_2$ gas have allowed us to study resonant nonlinear refraction for a range of normalized Rabi frequencies, $\upsilon_{Rabi} \Delta\upsilon^{-1}$, from 0.7 - 3. As described above, these effective Rabi frequencies were estimated as $\upsilon_{Rabi} = \mu E/h(\Delta\upsilon/\Delta\upsilon_L)^{1/2}$ where $\Delta\upsilon_L$ is the laser bandwidth. Figure 7 depicts the inferred sign of the resonant nonlinear refractive index as a function of $\upsilon_{Rabi} \Delta\upsilon^{-1}$ for the cases presented in sections III and IV. For simplicity, we have only plotted the sign of the nonlinearity for frequencies below resonance in Fig. 7. Remarkably and despite the large differences in pressure, dipole moment, and line spacing, the sign of the nonlinearity of both CO and CO$_2$ tends to reverse for $\upsilon_{Rabi} \Delta\upsilon^{-1} \sim 1$. The same trend in both CO and CO$_2$ suggests that the same mechanism is responsible for the optical nonlinearity of each molecular gas.

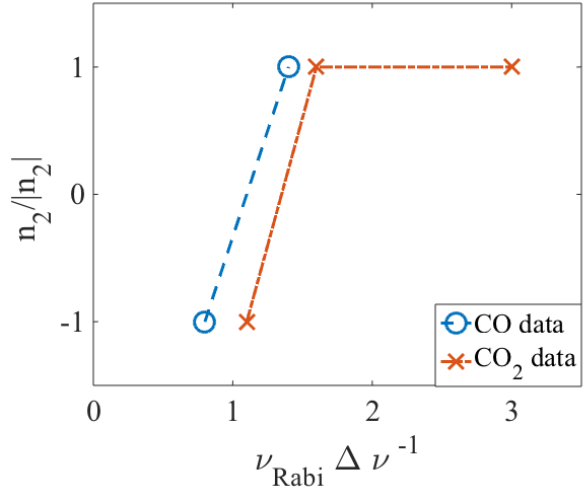

Figure 6: The sign of the nonlinearity inferred from CO and $CO_2$ data for frequencies below resonance plotted as a function of the estimated ratio between the effective Rabi frequency and the collisional linewidth.

## VI. Density matrix calculations for a strongly driven two-level system

Measurements of resonant nonlinear refraction in CO and $CO_2$ gas indicate that the nonlinearity of these simple molecules differs substantially from the resonant nonlinear optics of atomic gases and more complex molecules such as $SF_6$. The most striking difference is that the sign of the nonlinearity for both CO and $CO_2$ can change from positive to negative resulting in self-focusing or self-defocusing, respectively, depending on the ratio of the Rabi frequency to the collisional linewidth, $\upsilon_{Rabi}\,\Delta\upsilon^{-1}$. Moreover, time-resolved measurements of nonlinear refraction indicate that the nonlinear response time of these molecules is fast compared with the lifetime of the upper state. These observations suggest that field effects such as power-broadening and ac-Stark splitting that occur on a fast time scale dominate the nonlinearity. The purpose of this section is to provide a qualitative explanation of the physics behind these observations based on density matrix calculations of the electric susceptibility.

A full theoretical model of the resonant nonlinear optics of CO or $CO_2$ is rather complicated and requires substantial effort that goes beyond the scope of this experimental study. Due to the large bandwidth and intensity, the driving laser pulse may interact with more than one rotational-vibrational line at once. A sufficient model, even for CO, would therefore need to account for density matrix elements of two or more vibrational states along with a large number of rotational sub-levels for each vibrational level. This system of coupled differential equations would need to be self-consistently solved with a nonlinear wave equation describing the 3D evolution of the driving laser pulse that has a temporal pulse duration comparable with the dipole dephasing time. Ideally, such calculations should also resolve the carrier frequency of the laser pulse to account for the various nonlinear optical frequency mixing that may occur. Modeling the nonlinearity of $CO_2$ would prove to be an even more formidable task than that of CO since this molecule has multiple 4.3 µm bands that should be accounted for in a self-consistent model. Insight into the resonant nonlinear optics of CO and $CO_2$ may still be gleamed, however, through the analysis of a simplified two or three-level system model. It should be noted that steady-state, perturbative solutions of the equations of motion for the density matrix have been applied to problems ranging from optically pumped molecular lasers [30] to the resonant nonlinearity of atomic gases [*e.g.* 31, 32, 33] and complex solid-state systems [34].

The strong distortion of an absorption line that is characteristic of the ac-Stark effect can lead to sign reversals in the nonlinear refractive index as observed in this experiment. Such strong modulation of the absorption line is typically observed when exciting two transitions with a shared level in a cascade, Lambda, or V-type scheme [1, 30]. Since the vibrational levels of CO and $CO_2$ have a much smaller anharmonicity than electronic transitions, it is conceivable that the finite bandwidth of the pump laser could excite multiple transitions in a cascade-type configuration. However, a closer examination of the spectral location of the transitions of these molecules indicates that this is not achievable without a very large frequency detuning. Even an analysis of the least anharmonic molecule, $CO_2$, indicates that the regular band (000 – 001) and sequence band (001 – 002) 4.3 µm transitions that can be simultaneously excited do not have a shared energy level. Figure 8a is a plot of the absorption spectrum of $CO_2$ in the ~ 4.3 µm wavelength range calculated from HITRAN [25]. Figure 7a shows both the regular and sequence band transitions of $CO_2$ gas for a vibrational temperature of ~ 2500 K. It should be noted that previous vibrational temperature measurements used to study the gain dynamics of an optically pumped $CO_2$ laser [35] indicate that this temperature is reasonable for our experimental conditions. Figure 7b is a zoomed version of Fig. 7a with the 4P(34) line marked with an arrow as a reference. In Fig. 7b, we have also marked the sequence band transitions that have a lower energy level that corresponds to the upper energy level of the 4P(34) transition (*i.e.* 4R(33) and 4P(33)). Since the pump bandwidth used in experiment is approximately half as wide as the spacing between adjacent rotational lines, these transitions are too detuned from the pump frequency to be simultaneously excited. The same was true for the other transitions studied in this experiment.

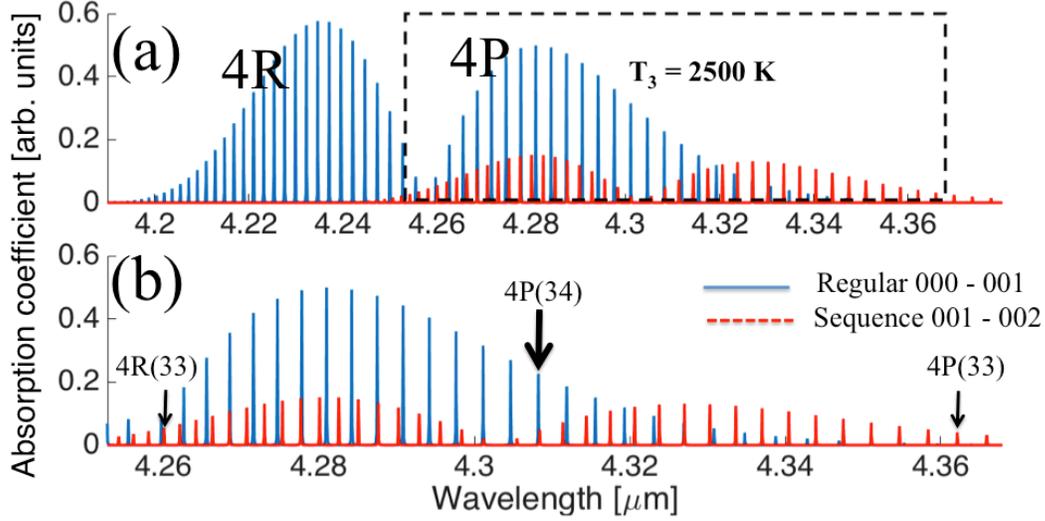

Figure 7: (a) Calculated absorption spectrum of $CO_2$ in the 4.3 μm wavelength region from HITRAN [25] and showing only the regular band transitions (000 – 001) and sequence band transitions (001 – 002). (b) Zoomed in the area of the black dashed rectangle in (a). The 4P(34) regular band transitions explicitly shares an energy level with the 4R(33) and 4P(33) sequence band transitions and are marked.

Since it is unlikely that the ac-Stark splitting associated with two coupled transitions in a three-level system can explain our observations, we have applied the density matrix description of a strongly driven, two-level system presented in [31]. This description, which involves the interaction of a strong pump wave and two weak probe waves, has been applied to describe the four-wave mixing of self-trapped filaments in Rb vapor [32]. Although we have not applied multiple laser fields to the molecular gases studied in this experiment, the relatively large laser bandwidth can provide the multiple frequencies required to illicit the nonlinear response described in [31]. Figure 8a, depicts the absorption spectrum of a single rovibrational line of $CO_2$ or CO with the approximate Fe:ZnSe pump laser bandwidth superimposed. Here we have denoted the strong pump's frequency by $\omega_0$ and the weak probe frequencies by $\omega_3$ and $\omega_4$. Figure 8b depicts the ac-Stark splitting of a strongly driven two-level system where the initial energy levels each split into two states that are separated by the generalized Rabi frequency, $\Omega' = (\Omega^2 + \Delta^2)^{1/2}$ where $\Omega$ is the Rabi frequency and $\Delta$ is the frequency detuning between the strong pump and the resonance frequency. As can be seen in Fig. 8b, the ac-Stark splitting provided by the pump generates new resonances that can affect the propagation of the probe waves.

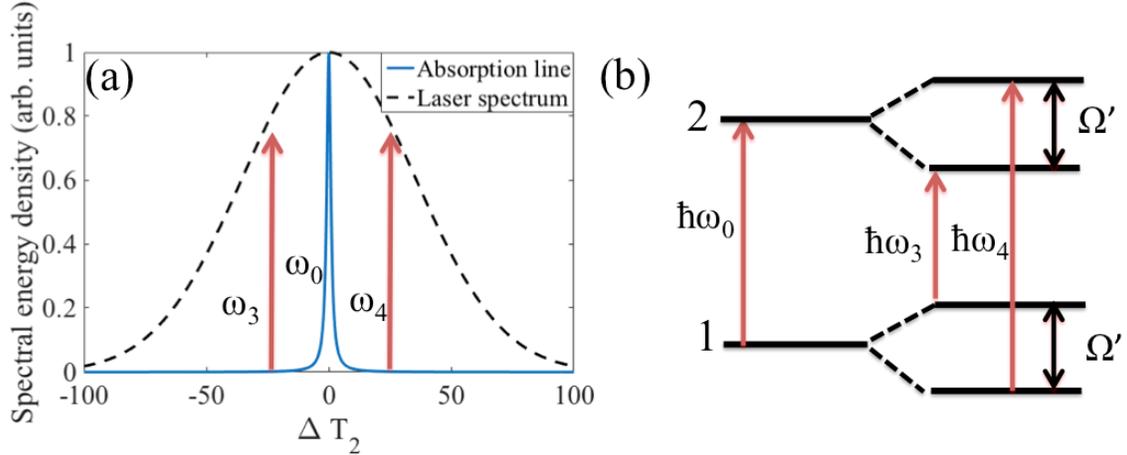

Figure 8: (a) Calculated Lorentzian absorption line of a single rovibrational line broadened to 100 torr with the approximate bandwidth of the Fe:ZnSe laser superimposed. (b) A depiction of the ac-Stark splitting of a two-level system with driven with a strong field, definitions of the frequencies are given in the text.

Several references [31, 36, 37] have presented steady-state, analytic solutions for the density matrix elements and related electric susceptibilities of the system depicted in Fig. 8b. These perturbative solutions for the density matrix are correct to all orders of the strong pump wave at frequency $\omega_0$ having an electric field amplitude $E_0$, and to first-order of the weak probe waves at frequencies $\omega_3$ and $\omega_4$ with electric field amplitudes denoted by $E_3$ and $E_4$, respectively [31]. The probe fields are assumed to be unable to saturate the molecular resonance without the presence of the pump wave. The solutions to the off-diagonal density matrix elements oscillating at $\omega_3$ are reproduced here for convenience:

$$\rho_{21}(\omega_3) = \frac{\hbar^{-1}\mu(\rho_{22}-\rho_{11})^{dc}}{D(\delta)} E_3 \left[\left(\delta + \frac{i}{T_1}\right)\left(\delta - \Delta + \frac{i}{T_2}\right) - \frac{1}{2}\Omega^2 \frac{\delta}{\Delta - \frac{i}{T_2}}\right], \quad (1a)$$

$$\rho_{21}(2\omega_0 - \omega_4) = \frac{2(\rho_{22}-\rho_{11})^{dc}\mu^3 E_0^2 E_4^* \frac{\left(-\delta-\Delta-\frac{i}{T_2}\right)\left(\delta+\frac{2i}{T_2}\right)}{\Delta+\frac{i}{T_2}}}{\hbar^3\left(\Delta+\delta+\frac{i}{T_2}\right)D^*(-\delta)}. \quad (1b)$$

Where $(\rho_{22} - \rho_{11})^{dc}$ is the steady-state saturated population inversion induced by the strong pump $E_0$:

$$(\rho_{22} - \rho_{11})^{dc} = (\rho_{22} - \rho_{11})^0 \frac{(1+\Delta^2 T_2^2)}{1+\Delta^2 T_2^2 + \Omega^2 T_1 T_2}, \quad (2)$$

and $(\rho_{22} - \rho_{11})^0$ is the population inversion at equilibrium. Finally, D(δ) is the cubic function that is given by,

$$D(\delta) = \left(\delta + \frac{i}{T_1}\right)\left(\delta - \Delta + \frac{i}{T_2}\right)\left(\delta + \Delta + \frac{i}{T_2}\right) - \Omega^2\left(\delta + \frac{i}{T_2}\right). \tag{3}$$

In the above equations, $T_1$ and $T_2$ are the upper state lifetime and the dipole dephasing time, respectively, and $\delta$ is the detuning of the probe waves from the pump wave. For simplicity, we have limited our calculations to those density matrix elements that oscillate at $\omega_3$, however, forms for the elements oscillating at $\omega_4$ are very similar. The off-diagonal density matrix elements can be cast in the form of an effective, intensity dependent, first- and third-order electric susceptibility using the following equations for the first and third-order polarizations, $P^{(1)}(\omega_3) = N\mu\rho_{21}(\omega_3) = \chi^{(1)}(\omega_3)E_3$ and $P^{(3)}(2\omega_0 - \omega_4) = N\mu\rho_{21}(2\omega_0 - \omega_4) = 3\chi^{(3)}(2\omega_0 - \omega_4)E_0^2 E_4^*$, where N is the number density of molecules. The effective susceptibilities are:

$$\chi^{(1)}(\omega_3) = \frac{N\mu^2(\rho_{22} - \rho_{11})^{dc}}{\hbar D(\delta)}\left[\left(\delta + \frac{i}{T_1}\right)\left(\delta - \Delta + \frac{i}{T_2}\right) - \frac{1}{2}\Omega^2 \frac{\delta}{\left(\Delta - \frac{i}{T_2}\right)}\right], \tag{4a}$$

$$\chi^{(3)}(2\omega_0 - \omega_4) = \frac{2N\mu^4(\rho_{22} - \rho_{11})^{dc}\frac{\left(-\delta - \Delta - \frac{i}{T_2}\right)\left(\delta + \frac{2i}{T_2}\right)}{\Delta + \frac{i}{T_2}}}{3\hbar^3\left(\Delta + \delta + \frac{i}{T_2}\right)D^*(-\delta)}. \tag{4b}$$

Let us consider the $CO_2$ molecule for calculations of the electric susceptibility since we have observed both self-focusing and self-defocusing in this gas. For 100 torr of $CO_2$ gas $T_1$ is $\sim 40$ µs [27] and $T_2$ is $\sim 0.6$ ns [38]. However, fast rotational relaxation acts to re-populate both the upper and lower levels on a $\sim 1$ ns time scale [39]. Since this re-population emerges from a rotational reservoir that is not accounted for, we have performed calculations with an effective upper state lifetime that is 1.7x ($\sim 1$ ns/0.6 ns) that of the dipole dephasing time. Calculations were performed assuming a maximum effective Rabi frequency of $\sim 1.5$ GHz that corresponds to a $\Omega T_2$ value in the $2 - 3$ range. It should be noted that $\Omega T_2 \sim \upsilon_{Rabi} \Delta\upsilon^{-1}$ for our experimental conditions. Finally, all calculations were carried out assuming that the pump wave is perfectly in resonance with the molecular transition frequencies, *i.e* $\Delta = 0$.

Figure 9 shows the calculated effective first-order susceptibility for the parameters described above. Figures 9a and 9c depict the imaginary and real parts of the susceptibility, respectively, for values of $\Omega T_2 < 0.5$ ($\Omega T_2 = 0.1$, 0.2, and 0.4) and Fig. 9b and 9d show the imaginary and real parts of the susceptibility for $\Omega T_2 \geq 0.5$ ($\Omega T_2 = 0.5$, 1, and 2). As can be inferred from Fig. 9, the susceptibility transitions from the saturable absorption regime (see Fig. 9a and 9b) to produce an absorption line that is split or modulated due to the strong pump field. For these parameters, the transition occurs for a $\Omega T_2$ parameter near unity but can change depending on the ratio $T_1/T_2$ [31]. The inset of Fig. 9d is a zoomed version of the same plot close to the line center and depicts the beginning of a sign reversal in the change of the first-order susceptibility as a function of laser intensity – *e.g.* red photons experience a larger susceptibility at $\Omega T_2 = 2$ than at $\Omega T_2 = 1$.

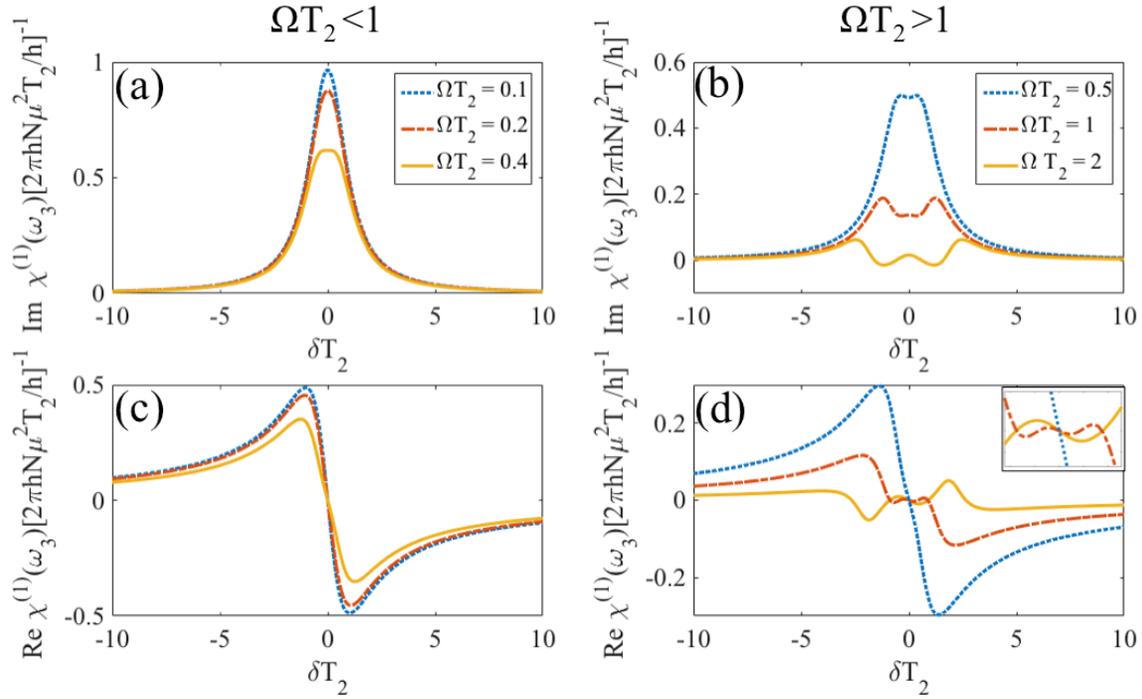

Figure 9: The imaginary part of the first-order electric susceptibility for Rabi frequencies (a) less than the collisional linewidth and (b) greater than the collisional linewidth. The real part of the first-order electric susceptibility for Rabi frequencies (c) less than the collisional linewidth and (b) greater than the collisional linewidth.

The third-order nonlinearity must also be considered to calculate the total susceptibility acting on the laser pulse. Figure 10a and 10b are plots of the real part of the third-order nonlinearity for $\Omega T_2 < 0.5$ and $\Omega T_2 \geq 0.5$, respectively, calculated for the same values of $\Omega T_2$ that were used to generate the results presented in Fig. 9. As can be seen in Fig. 10a and 10b, the third-order nonlinearity grows with increased field strength until $\Omega T_2$ reaches a value near unity and then begins to decrease in amplitude and change shape. Figure 10c and 10d are plots the total susceptibility $\chi_{total} = \chi^{(1)}(\omega_3) + 3\chi^{(3)}(2\omega_0 - \omega_4)|E_0|^2$, for $\Omega T_2 < 0.5$ and $\Omega T_2 \geq 0.5$, respectively. These results represent a sum of the first-order susceptibilities plotted in Fig. 9c and 9d with the third-order susceptibilities plotted in Fig. 10a and 10b. The total susceptibility plotted in Fig. 10c for $\Omega T_2 < 0.5$ is similar to a nonlinearity that is dominated by saturable absorption in that red and blue photons are subjected to a self-defocusing and self-focusing nonlinearity, respectively. The total electric susceptibility for the strongly driven system (see Fig. 10d) shows, however, that the nonlinearity for frequencies within a few collisional linewidths of the line center (shaded region of Fig. 10d) is similar to the weakly driven case (Fig. 10c) for $\Omega T_2$ between 0.5 and 1 and begins to change sign when driven at $\Omega T_2 > 1$. These calculations indicate that, for a broadband laser pulse, the nonlinearity can change sign for Rabi frequencies comparable to the collisional linewidth. This finding is qualitatively consistent with the experimental observation of intensity-dependent sign reversals for in the nonlinear refractive index of both CO and $CO_2$ for Rabi frequencies greater than or equal to the collisional linewidth ($\upsilon_{Rabi} \Delta\upsilon^{-1} \sim \Omega T_2 > 1$).

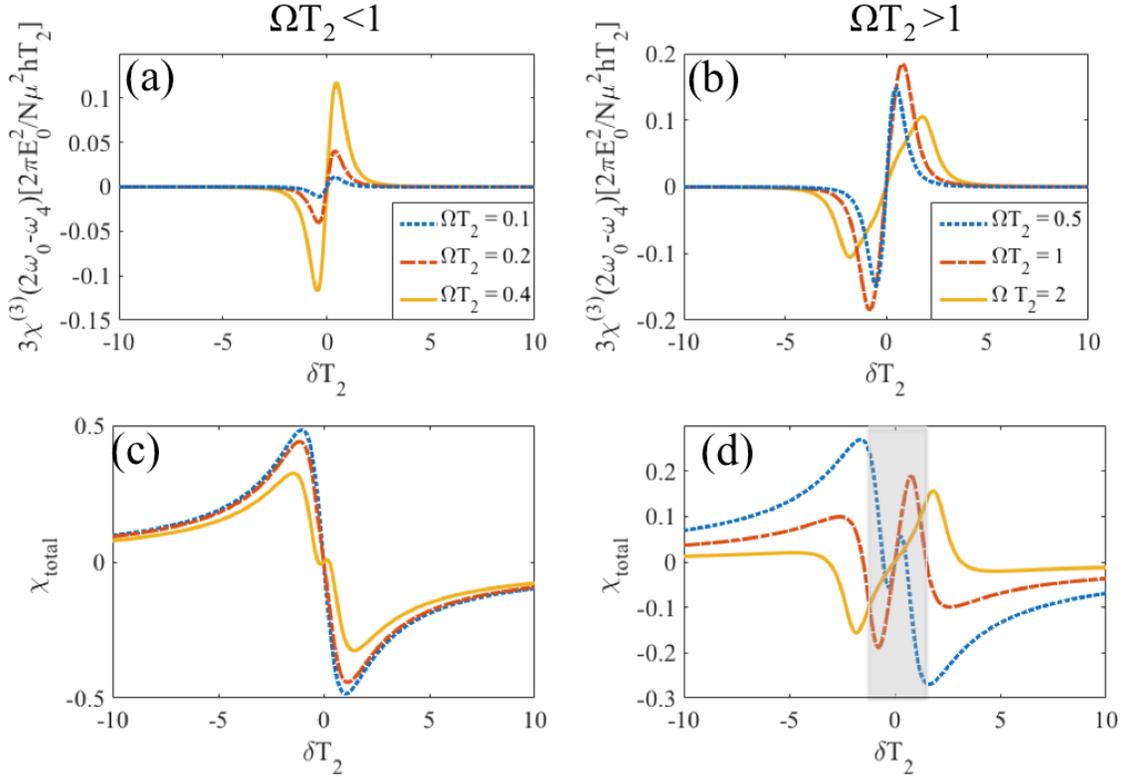

Figure 10: The real part of the third-order electric susceptibility for Rabi frequenices (a) less than the collisional linewidth and (b) greater than the collisional linewidth. The real part of the total electric susceptibility (sum of results presented in (a) and (b) with those presented in Fig. 9c and 9d) for Rabi frequencies (c) less than the collisional linewidth and (b) greater than the collisional linewidth.

## VII. Discussion

The main finding of this experimental study was that the resonant nonlinearity of CO and $CO_2$ could change from self-focusing to self-defocusing and vice-versa depending on the intensity of the driving laser pulse. Frequency-resolved nonlinear refraction measurements obtained for each molecular gas, indicate that the ratio of the Rabi frequency to the collisional linewidth ($\upsilon_{Rabi}\, \Delta\upsilon^{-1}$) is the critical parameter for determining the sign of the resonant nonlinearity in this regime. This notion is clearly evident from Fig. 6, above, that summarizes the frequency-resolved nonlinear refraction data by depicting the sign of the resonant nonlinearity as a function of the normalized Rabi frequency, $\upsilon_{Rabi}\, \Delta\upsilon^{-1}$. As can be inferred from Fig. 6, these molecules exhibit a sign of nonlinearity consistent with a system dominated by saturable absorption [7] for low intensities ($\upsilon_{Rabi}\, \Delta\upsilon^{-1} < 1$) and exhibit the opposite sign of nonlinearity for high intensities ($\upsilon_{Rabi}\, \Delta\upsilon^{-1} > 1$).

Nonlinear refractive index sign reversals occurring for Rabi frequencies larger than the collisional linewidth are also predicted by density matrix calculations of the total susceptibility of a strongly driven two-level system [31]. As can be inferred by comparing the high intensity ($\Omega T_2 \sim \upsilon_{Rabi}\, \Delta\upsilon^{-1} < 1$) and low intensity ($\Omega T_2 \sim \upsilon_{Rabi}\, \Delta\upsilon^{-1} > 1$) cases depicted in Fig. 10a and Fig. 10b, respectively, the density matrix model predicts that a sign reversal in the susceptibility will occur for Rabi frequencies that become larger than the collisional linewidth. The qualitative consistency between our experimental observations and the simplified theory suggests that the nonlinearity of CO and $CO_2$ can be understood as a strongly driven two-level system. In this model, the resonant nonlinearity changes sign due to population oscillations that occur at beat frequencies provided by the large bandwidth of the pump pulse. These population oscillations act to split or modulate the absorption line similar to the ac-Stark effect [31] that, in turn, cause the sign of the nonlinearity to change. The model also suggests that the response time of such a nonlinearity will be on order of the dipole dephasing time, ~ 0.6 ns for 100 torr of $CO_2$ [38], which is consistent with the < 4 ns response time determined by time-resolved measurements. Finally, the observation of absorption features in the frequency-resolved data are also consistent with this physical picture and suggest that field effects strongly contribute to the observed optical nonlinearity of these molecules (see Fig. 3).

Although this model provides a reasonable explanation for these observations, it is oversimplified in a number of significant ways and thereby falls short in providing a complete description of resonant nonlinear refraction of mid-IR light in CO and $CO_2$ gas. Most notably, the strongly driven two level system model predicts that the molecular resonance saturates for such large intensities while no sign of saturation was observed in the experiment. The rotational structure of CO and $CO_2$ that is unaccounted for in the two-level model may explain this discrepancy. Indeed, previous studies have indicated that similar simple molecules such as $SO_2$, OCS, $NO_2$, and $O_3$ will not exhibit saturation even at intensities of $10^9 - 10^{11}$ W/cm$^2$ [23, 24].

## VIII. Conclusion

We have used frequency- and time-resolved measurements of resonant nonlinear refraction of mid-IR light in CO and $CO_2$ gas to study the resonant nonlinear optics of these simple molecules. Our measurements and density matrix calculations of a strongly driven two-level system indicate that field effects such as power-broadening and ac-Stark splitting dominate the nonlinear response resulting in intensity dependent, nonlinear refractive index sign reversals. For each molecular gas, the sign reversal in the optical nonlinearity was observed to occur for Rabi frequencies that are comparable or larger than the collisional linewidth. These measurements indicate that, for relatively narrowband lasers, the critical power for self-focusing of mid-IR light in the air can be reduced by up to $10^4$ times by IR-active, minor air constituents. This large and fast nonlinearity may be used to enable the filamentation of joule-class, 0.01 – 1 ns, mid-IR pulses in the air. These measurements and similar ones for other IR-active minor air constituents may be used to develop a nonlinear refractive index equivalent to the widely used HITRAN molecular database.

**Acknowledgements**

This material is based upon work supported by the Office of Naval Research (ONR) MURI (N00014-17-1-2705) and the Department of Energy (DOE) Office of Science Accelerator Stewardship Award No. DE-SC0018378.